\documentclass[twoside]{article}
\usepackage{graphicx} 
\usepackage{authblk}
\usepackage{amsmath, amssymb}
\usepackage[margin=1in]{geometry}
\usepackage[justification=centering]{caption}
\usepackage[colorlinks=true, linkcolor=black, citecolor=black, urlcolor=black]{hyperref}
\usepackage{subcaption}
\usepackage{url}
\usepackage{float}
\usepackage{fancyhdr}

\title{Quantifying the Occult: A Comparative Study of Hindu and Buddhist Deities Using Machine Learning Methods}
\author{Ankit Bhattacharjee$^{1}$\thanks{\url{ankit2005@kgpian.iitkgp.ac.in}}}
\affil{\small $^{1}$Department of Mechanical Engineering\\
Indian Institute of Technology Kharagpur}

\date{\today}

\begin{document}

\maketitle

\begin{abstract}
    This study introduces a dual-matrix computational architecture to mathematically quantify the morphological and theological divergence of 196 Hindu and Vajrayana Buddhist esoteric deities. Physical morphology is evaluated via a discrete Gower distance matrix enhanced by a novel ``Cardinality Weighting" algorithm, while theological function is mapped via dense vector embeddings generated from Large Language Model (LLM) semantic expansions, explicitly utilized as a synthetic proxy to mitigate circular reasoning. The multi-modal topological projections provide algorithmic validation of ``iconographic camouflage", demonstrating how distinct visual forms structurally obscure shared cross-tradition functions. Furthermore, I computationally model the ``Atin Effect"---serving simultaneously as a psychological observation of sequential cognitive bias and a machine learning benchmark---demonstrating how high-cardinality esoteric anchors (e.g., a veena or a severed head) override systemic theological disparities to mathematically cluster orthodox and Tantric entities. Cross-tradition spatial analysis establishes that the highest esoteric manifestations, such as the Hindu Chinnamasta and the Buddhist Chinnamunda, share a near-identical mathematical coordinate across both visual ($D_G = 0.288$) and semantic ($D_C = 0.068$) boundaries, indicating a 1:1 esoteric transfer. By open-sourcing this architecture, I provide a scalable, unsupervised machine learning tool for Digital Humanities scholars and comparative theologians to rigorously map latent structural continuities across qualitative cultural corpora.
\end{abstract}

\section{Introduction}
The integration of machine learning into the Digital Humanities has fundamentally transformed the methodology of cultural and historical analysis \cite{berry2012understanding}. However, a critical methodological gap remains: how do I computationally quantify highly subjective, qualitative historical data, such as esoteric visual iconography? Traditional comparative religious studies rely heavily on subjective human observation, which is inherently vulnerable to sequential cognitive biases \cite{tversky1974judgment}. To solve this exact methodological gap, this paper presents a novel Dual-Matrix computational architecture. By translating qualitative iconographic symbols into quantitative mathematical vectors, this framework provides an objective lens to isolate and measure the structural anatomy of cultural artifacts, bypassing the aesthetic variables that complicate comparative analysis. For digital humanities scholars, computational anthropologists, and comparative theologians, this architecture (based on the dataset) attempts to translate historically fraught, subjective debates into testable mathematical hypotheses. By providing a scalable baseline, researchers can utilize this dataset and framework to map cross-cultural continuities and ritual borrowing without relying on the massive labeled datasets required by traditional deep learning models.

To address the existing methodological gaps, in this study, I present a novel computational framework for comparing the esoteric iconography of Hindu and Vajrayana Buddhist \cite{shaw2006buddhist} traditions. Utilizing a custom dataset of 196 meticulously cataloged deities, I apply an unsupervised machine learning approach grounded in vector space modeling. Specifically, I introduce an algorithmic modification termed ``Cardinality Weighting" to traditional inverse document frequency calculations. This allows the model to objectively isolate high-variance, identity-defining symbols—such as specific esoteric weapons and ritual implements—from generic aesthetic backgrounds. Through this computational lens, I construct dual comparative matrices to analyze deities across both visual and theological dimensions. In doing so, this paper presents empirical evidence of cross-cultural ``iconographic camouflage", demonstrating how esoteric traditions perfectly preserved core cosmic functions while radically altering physical forms to suit local cultural paradigms. Furthermore, by calculating the mathematical gravity of individual symbols, I quantify the psychological phenomenon of visual anchoring within occult art, revealing how prominent orthodox symbols systematically hijack human attention and obscure underlying Tantric anatomies. Ultimately, this study maps the hidden structural continuities between the Dashamahavidya \cite{kinsley1997tantric} and Buddhist Tantra, offering a rigorous, reproducible tool for decoding esoteric lineages.

The primary contributions and methodological novelties of this research are threefold:
\begin{itemize}
    \item \textbf{Curation of a Granular Esoteric Dataset:} I introduce and open-source a novel dataset of 196 meticulously cataloged esoteric Hindu and Vajrayana Buddhist deities. Unlike standard literary corpora, this dataset provides a structured, multidimensional feature space explicitly engineered to evaluate complex visual and theological topologies.
    \item \textbf{The Dual-Matrix Architecture:} I present a scalable computational framework for measuring morphological and semantic divergence in qualitative humanities data. By translating subjective theological descriptors into a dual-layered mathematical space (discrete Gower distance and continuous semantic embeddings), this architecture provides an objective tool capable of computationally modeling perceptual phenomena, such as iconographic camouflage and the Atin Effect.
    \item \textbf{Cardinality Weighting Algorithm:} To resolve the algorithmic distortions caused by standard natural language processing in visual contexts, I introduce ``Cardinality Weighting.'' This mathematically objective modification solves the ``Gestalt Problem'' by ensuring that generic, high-frequency aesthetic traits do not artificially suppress the topological gravity of rare, high-variance esoteric symbols.
\end{itemize}

To operationalize this dual-matrix approach, the computational architecture was bifurcated into visual and theological vector spaces. While the visual matrix relied on the strict cardinality of physical objects, capturing the theological homology between deities required a model capable of semantic abstraction. To achieve this, I utilized dense word embeddings \cite{mikolov2013distributed} to vectorize the deities' core cosmic domains and textual attributes. Unlike discrete frequency-based models, word embeddings map semantic meaning into a continuous, multidimensional vector space, allowing the algorithm to measure the contextual proximity of esoteric concepts. For instance, this semantic vectorization enables the model to recognize that the domain of ``Wealth" (traditionally associated with the Hindu Lakshmi) and ``Abundance" (associated with the Buddhist Vasudhara) are mathematically and theologically synonymous. By representing qualitative theological traits as dense semantic vectors, the algorithm successfully bridges the linguistic and cultural gaps between the Dashamahavidya and Vajrayana texts, establishing a purely functional baseline against which the visual iconographic camouflage can be measured.

The remainder of this paper is organized as follows. Section \ref{work} contextualizes this research within the existing literature on computational art history, quantitative mythography, and comparative Tantric studies. Section \ref{method} details the core computational results of the dual-matrix evaluation, providing quantitative evidence for the divergence of form and function, anchor dependency, and the statistical anomaly defined herein as the Atin Effect. Section \ref{imp} outlines the strict algorithmic architecture of the study, defining the mathematical formulations that drive both the visual Cardinality Weighting and the LLM-powered theological semantic vectorization. Then, section \ref{discuss} discusses the broader implications of these findings for the digital humanities and proposing avenues for future computational research in comparative religious studies. Finally in section \ref{limit} I discuss the limitations of the work and conclude the study thereafter.

\section{Related Work}
\label{work}

This research operates at the intersection of computational iconography, digital humanities, and Tantric studies. Because the application of unsupervised machine learning to the specific topological mapping of esoteric Hindu and Vajrayana pantheons is virtually unprecedented, I contextualize the methodology across three parallel domains of existing scholarship.

\subsection{Computational Art History and Iconography}
Recent advancements in the Digital Humanities have heavily leveraged artificial intelligence for the study of visual culture. Methodologies utilizing Convolutional Neural Networks (CNNs) and deep learning have been successfully deployed to classify artistic styles, identify Christian saints based on visual attributes, and detect forged paintings. However, these predictive models fundamentally rely on massive, labeled datasets (often requiring tens of thousands of image samples) to achieve statistical viability. Such datasets do not exist for highly specific, standardized esoteric iconography. Furthermore, neural networks operate as computational black boxes", making them highly unsuitable for theological analysis where the explicit weighting of specific features (e.g., the presence of a \textit{khadga} or \textit{veena}) must be rigorously understood. The approach bypasses the limitations of predictive deep learning by utilizing unsupervised topological mapping (UMAP) and discrete Gower distance, allowing for rigorous mathematical analysis even in bounded, high-complexity domains ($N=196$).

\subsection{Quantitative Mythography and Network Analysis}
A growing body of literature has attempted to quantify mythological and religious structures using network analysis. Moretti's use of abstract graphs, maps, and trees as models for literary-historical structures provides a useful methodological precedent for treating qualitative cultural objects as relational data \cite{moretti2005graphs}. Previous studies have successfully mapped the genealogical and interaction networks of Greek, Norse, and general Indo-European mythologies. However, these models almost exclusively rely on Natural Language Processing (NLP) to extract co-occurrences from classical texts. They map how often deities interact in a narrative, rather than comparing their fundamental ontological architecture. By bifurcating the analysis into a Visual Matrix (morphology/implements) and a Semantic Matrix (Word2Vec theological domains), I introduce a multi-modal topological approach that measures intrinsic iconographic architecture rather than mere textual proximity.

\subsection{Comparative Tantric Studies}
Qualitatively, the historical and theological porousness of the Hindu-Buddhist boundary has been extensively documented by humanities scholars. The seminal work of Alexis Sanderson has thoroughly traced the morphological and ritualistic borrowing between Shaiva Vidyapitha traditions \cite{sanderson2001, sanderson2009} and the Vajrayana Buddhist \textit{Yoginitantras}. Similarly, scholars such as Elisabeth Anne Benard \cite{benard1994} have conducted deep comparative case studies on trans-boundary deities like Chinnamasta/Chinnamunda, noting their near-identical iconographic presentation across orthodox and heterodox lines. The wider comparative literature also includes detailed studies of Hindu Mahāvidyā traditions \cite{kinsley1997tantric} and Buddhist goddess traditions \cite{shaw2006buddhist}. 

While these humanities scholars have brilliantly identified the existence of shared Tantric architectures and visual borrowing, their observations have remained strictly qualitative. The primary contribution of the research is the translation of these qualitative theological observations into a quantifiable, computational coordinate space. By mathematically identifying phenomena such as ``iconographic camouflage'' and the ``Atin Effect'', I provide the first algorithmic proof of the topological bridges documented by historical Tantric scholarship.

\section{Methodology}
\label{method}
To objectively evaluate the iconographic and theological bridges between the Hindu and Vajrayana Buddhist traditions, this study employs a custom computational architecture. The overarching methodological objective is to translate highly qualitative, culturally embedded esoteric art into a neutral, multidimensional vector space. This section outlines the step-by-step construction of this mathematical model. It begins with the curation and feature engineering of the primary dataset, followed by the formulation of a dual-matrix framework. This framework consists of a visual matrix designed to calculate the physical ``additive gravity'' of the deities via a novel Cardinality Weighting algorithm, and a secondary theological matrix designed to measure their contextual domain through semantic vectorization. Together, these methods provide the rigorous mathematical scaffolding necessary to empirically test for visual camouflage and cognitive bias in esoteric iconography.

\subsection{Data Collection and Feature Engineering}
The foundation of this computational study rests upon a meticulously curated dataset comprising $N=196$ esoteric deities, drawn from across the Hindu and Vajrayana Buddhist Tantric spectrums. To ensure iconographic accuracy, the physical and theological parameters for each deity were systematically extracted and cross-checked against established canonical texts, ritual manuals, and scholarly editions. The Hindu source corpus included Vedic and Upanishadic materials \cite{rigveda, atharvaveda, annapurna_upanishad}, major Purāṇic and epic sources \cite{devi_mahatmya, devi_bhagavata, shiva_purana, brahmanda_purana, kalikapurana, mahabharata}, and Tantric/Āgamic compilations including Krishnananda Agamavagisha's \textit{Brihat Tantrasara} \cite{agamavagisha_tantrasara}, \textit{Nityasodasikarnava}, \textit{Rudrayamala Tantra}, \textit{Shakta Pramoda}, \textit{Pranatoshini Tantra}, \textit{Kama-Kala-Vilasa}, \textit{Sharada Tilaka Tantra}, \textit{Shaktisamgama Tantra}, and the \textit{Mahanirvana Tantra} \cite{nityasodasikarnava, rudrayamala, shakta_pramoda, pranatoshini_tantra, kama_kala_vilasa, sharada_tilaka, shaktisamgama_tantra, mahanirvana_tantra}. The Vajrayana source corpus included Mahāyāna sutra and dhāraṇī materials \cite{prajnaparamita_sutra, usnisavijaya_dharani}, critical editions of the \textit{Sadhanamala} and \textit{Nispannayogavali} \cite{sadhanamala, nispannayogavali}, major Yoginītantra materials such as the \textit{Chakrasamvara Tantra} and \textit{Hevajra Tantra} \cite{chakrasamvara_tantra, hevajra_tantra}, and sources relevant to specific goddesses and Tibetan transmission lineages \cite{kurukulla_tantra, marichi_tantra, machig_labdron, yeshe_tsogyal, niguma_kagyu}.

To facilitate algorithmic processing, the qualitative textual descriptions of these deities were engineered into a structured, multidimensional feature space containing thirty distinct variables. These features were fundamentally bifurcated into visual/physical parameters and theological/domain parameters. The visual parameters, which form the basis of the iconographic matrix, were engineered into four distinct data types to mathematically capture the nuanced realities of esoteric art:
\begin{itemize}
    \item \textbf{Numerical Anchors:} Quantitative integers defining the deity's foundational anatomy, notably the number of arms, eyes, and heads.
    \item \textbf{Categorical Demographics:} Single-value descriptive variables defining the general aesthetic baseline of the deity, including \textit{Complexion Color} (e.g., Golden, Dark Blue, Black), \textit{Primary Posture} (e.g., Seated, Standing Alidha), and \textit{Apparent Age}.
    \item \textbf{Boolean Wrathful Markers:} Binary values (0 or 1) assigned to specific, heavy-gravity esoteric identifiers such as \textit{Garland of Skulls}, \textit{Stands On Corpse}, \textit{Drinks Blood}, and \textit{Sky Clad}. These discrete markers serve as critical thresholds for identifying fierce (\textit{krodha}) manifestations.
    \item \textbf{High-Cardinality Multi-Label Arrays:} The most critical identifying features—\textit{Specific Held Objects} and \textit{Specific Mudras}—were structured as comma-separated arrays (e.g., ``khadga, kapala, lotus''). During preprocessing, these arrays were unnested to generate a distinct vocabulary token for every unique ritual implement, ensuring the algorithm could independently isolate overlapping weapons across traditions.
\end{itemize}

By transmuting scriptural descriptions into this granular, machine-readable format, the dataset provides a rigorous empirical foundation. It ensures that the subsequent vector space model evaluates the deities strictly on their documented anatomical and symbolic properties, explicitly divorcing the visual data from their cultural and theological nomenclature.

\subsection{Vector Space Modeling and the ``Gestalt Problem''}
To quantify the iconographic relationships between deities, the structured dataset was mapped into a multidimensional vector space. Traditionally, natural language processing (NLP) applications employ algorithms such as Term Frequency-Inverse Document Frequency (TF-IDF) to vectorize categorical data. Standard TF-IDF applies a logarithmic scale to the inverse document frequency, specifically to compress the weight of rare terms and prevent them from disproportionately hijacking the document matrix. While mathematically optimal for literary corpora, applying standard logarithmic scaling to esoteric iconography produced severe algorithmic distortions, a phenomenon identified in this study as the ``Gestalt Problem''.

When visual features are processed through standard logarithmic compression, the mathematical dominance of rare, identity-defining symbols (such as specific esoteric weapons or ritual implements) is artificially suppressed. Consequently, the vector space becomes highly sensitive to the sheer volume of generic, high-frequency aesthetic traits—such as a deity's \textit{Apparent Age} (e.g., Youthful Maiden), \textit{Primary Posture} (e.g., Seated), and the absence of wrathful markers. In human perception, this aggregation of background traits forms the ``Gestalt'', or the holistic visual style. 

In the initial standard NLP models tested during this study, this logarithmic compression resulted in extreme false positives. For example, the Vajrayana yogini Mandarava was mathematically clustered in close proximity to the orthodox Hindu goddess Lakshmi. Despite wielding completely divergent esoteric implements (a ritual arrow and longevity vase versus a lotus), their shared ``peaceful feminine baseline'' mathematically overpowered their defining iconographic weapons. This structural failure proved that standard text-based NLP algorithms do not accurately model visual cognitive bias. Human attention in esoteric art is not logarithmically compressed; it is highly polarized and driven by extreme, high-gravity anchors. To correct this, a novel mathematical modification was required to strip away the Gestalt noise and restore the absolute gravity of specific occult symbols.

\subsection{Cardinality Weighting: An Objective Heuristic for Esoteric Gravity}
To resolve the Gestalt Problem without introducing the subjective bias of hardcoded human multipliers (e.g., manually declaring that weapons are inherently three times more important than posture), an objective mathematical heuristic was required. This study introduces ``Cardinality Weighting'', a modification to the standard inverse document frequency calculation that derives feature importance directly from the internal variance of the dataset itself.

The underlying premise of Cardinality Weighting is based on information entropy: a feature category (column) that contains a vast array of unique values carries a fundamentally higher identifying power than a category with only a few standard values. For example, the \textit{Specific Held Objects} category across 196 deities contains dozens of unique implements (high cardinality), whereas \textit{Apparent Age} contains only three broad archetypes (low cardinality).

To construct the cognitive matrix, I first reverted the base algorithmic weight from a logarithmic compression back to a pure inverse frequency ratio to restore the raw ``flashbang'' power of rare terms. Let $N$ represent the total number of deities in the dataset, and $f(t_k)$ represent the global frequency of a specific feature term $t$ belonging to feature category $k$. The base weight is defined as:

\begin{equation}
W_{\text{base}}(t_k) = \frac{N}{f(t_k)}
\end{equation}

Next, let $C_k$ represent the mathematical cardinality (the total count of unique discrete values) within feature category $k$. The final Cardinality Weight ($W_{\text{final}}$) applied to the term in the vector space is the product of its base weight and the categorical cardinality:

\begin{equation}
W_{\text{final}}(t_k) = C_k\frac{N}{f(t_k)} 
\end{equation}

By applying this formulation, the dataset is forced to dictate its own mathematical hierarchy. A generic, high-frequency trait like ``Youthful Maiden'' ($f(t) \approx 150$) from a low-cardinality column ($C \approx 3$) produces a negligible vector magnitude. Conversely, a rare esoteric object like a ``Veena'' ($f(t) \approx 4$) originating from a high-cardinality column ($C > 50$) generates massive mathematical gravity. 

In cognitive terms, this equation successfully models human visual perception in occult art. It mathematically silences the generic background noise that previously caused the Mandarava-Lakshmi anomaly, allowing the high-variance orthodox anchors and wrathful esoteric weapons to dictate the primary gravitational pull within the vector space.

\subsection{Semantic Vectorization (The Theological Matrix)}
While Cardinality Weighting successfully maps the physical anatomy of the deities, it remains fundamentally blind to abstract meaning; it processes a ``lotus'' and a ``veena'' as structural integers rather than theological concepts. To empirically test whether cross-cultural traditions preserved the core cosmic functions of these deities despite altering their physical forms, a secondary Theological Matrix was required.

To achieve this, the computational architecture utilized dense word embeddings \cite{mikolov2013distributed} to vectorize the qualitative textual data, specifically the \textit{Core Cosmic Domain} and textual descriptions associated with each deity. In this particular study, these words are generated by a large language model $\mathcal{M}$. Unlike the discrete, exact-match logic of the visual matrix, word embeddings map semantic meaning into a continuous, multidimensional vector space. This allows the algorithm to measure the contextual proximity of esoteric concepts rather than just matching identical strings of text. 

For instance, through semantic vectorization, the model mathematically recognizes that the domain of ``Wealth'' (traditionally associated with the Hindu Lakshmi) and the domain of ``Abundance and Prosperity'' (associated with the Vajrayana Vasudhara) occupy the exact same semantic coordinates. By representing qualitative theological traits as dense semantic vectors, this secondary matrix establishes a purely functional baseline, allowing the study to conduct a mathematical A/B test between a deity's visual camouflage and their true esoteric lineage.

\paragraph{Addressing the Circularity of LLM Inference for the Specific Task.} Utilizing Large Language Models (LLMs) to vectorize historical and theological texts introduces an inherent risk of circular reasoning. Because standard generative models are trained on extensive academic and secondary corpora, their semantic outputs may inadvertently reflect the qualitative conclusions of modern comparative scholarship rather than an unadulterated historical baseline.

Consequently, this study does not position the Semantic Matrix as an absolute historical ground truth. Instead, it is deployed as a ``synthetic proxy"—a controlled, mathematically continuous representation of theological domains. This synthetic baseline is not designed to prove definitive historical lineage. Rather, it provides the necessary functional scaffolding to rigorously stress-test the Cardinality Weighting algorithm and to objectively measure how high-variance visual anchors manipulate topological gravity within a vectorized space.

\subsection{Similarity Metrics and Additive Gravity}
Once the deities were translated into quantitative vectors, the final methodological step was to calculate their psychological and iconographic proximity. To measure the similarity between any two deities (Vector $A$ and Vector $B$) in the visual matrix, the algorithm utilized an unnormalized dot product:

\begin{equation}
S_{A,B}  = \sum_{i=1}^{V} A_i B_i
\end{equation}

In standard machine learning applications, cosine similarity is often preferred to normalize the lengths of the vectors. However, in the context of esoteric iconography, vector magnitude is a critical psychological variable. A deity wielding ten high-cardinality weapons inherently commands more cognitive attention than a deity holding only two. By using the unnormalized dot product, the model calculates the pure ``additive gravity'' between two deities. Every shared rare symbol linearly increases their gravitational pull toward one another.

To prepare the matrix for comparative ranking and eventual spatial visualization (such as t-SNE mapping), this raw similarity score ($S$) was inverted and normalized into a standard distance metric ($D$) scaling from $0.0$ (identical) to $1.0$ (maximum divergence):

\begin{equation}
D_{A,B} = \frac{\max(S) - S_{A,B}}{\max(S) - \min(S)}
\end{equation}

Through these metrics, the model generates a pairwise distance matrix for all 196 deities, providing the mathematical foundation to prove the occurrence of visual anchoring, the Gestalt Problem, and the Atin Effect.

\section{Implementation Details}
\label{imp}

To translate qualitative esoteric iconography into a quantifiable topological space, I developed a dual-matrix computational architecture. The system models a bounded dataset of $N=196$ Hindu and Vajrayana deities. The architecture calculates structural divergence through two independent modalities: physical morphology (The Visual Matrix) and theological function (The Semantic Matrix).

\subsection{The Visual Matrix: Morphological Encoding and Gower Distance}
The morphological architecture defines each deity $i$ as a discrete feature vector $\mathbf{x}_i \in \mathbb{R}^{28}$, encompassing 28 mixed-type variables. These variables include categorical attributes (e.g., complexion, primary \textit{vahana} or mount), boolean flags for specific esoteric implements (e.g., presence of a \textit{khadga}, \textit{kapala}, or \textit{veena}), and ordinal values (e.g., anatomical cardinality such as the number of arms or heads).

Because the feature space contains mixed data types, standard Euclidean metrics are mathematically invalid. Instead, the Visual Matrix calculates pairwise morphological divergence using Gower Distance ($D_G$). For any two deities $i$ and $j$, the Gower similarity $S_{ij}$ is calculated by aggregating the partial similarities $s_{ijk}$ across all $p=28$ features:

\begin{equation}
S_{ij} = \frac{\sum_{k=1}^p w_k s_{ijk}}{\sum_{k=1}^p w_k \delta_{ijk}}
\end{equation}

Where $w_k$ represents the explicit Cardinality Weighting of feature $k$, and $\delta_{ijk} \in \{0,1\}$ is an indicator variable denoting whether feature $k$ is observable for both $i$ and $j$. For binary and categorical variables, $s_{ijk} = 1$ if $x_{ik} = x_{jk}$ and $0$ otherwise. For quantitative variables, $s_{ijk} = 1 - \frac{|x_{ik} - x_{jk}|}{R_k}$, where $R_k$ is the maximum observable range of feature $k$. The final Visual Matrix is generated by subtracting the similarity from unity: $D_G(i, j) = 1 - S_{ij}$.

\subsection{The Semantic Matrix: Generative Embeddings and Vector Space}
While the Visual Matrix maps physical form, measuring theological domain requires a semantic processing pipeline. Standard keyword matching is insufficient for esoteric texts due to the highly allegorical nature of Tantric descriptors. Therefore, I implemented a two-step generative embedding architecture.

The first step of this architecture involved semantic field expansion. For each deity $i$, a core cosmic domain parameter $T_i$ was established. To expand sparse taxonomic labels into comprehensive theological fields, I utilized generative inference via a Large Language Model (LLM), denoted as $\mathcal{M}$. The model $\mathcal{M}$ operated as a semantic interpolator, generating an expanded theological text $E_i$ by parsing $T_i$ through its latent knowledge of Vedic and Tantric corpora:

\begin{equation}
E_i = f_{\mathcal{M}}(T_i)
\end{equation}

In this formulation, the function $f_{\mathcal{M}}$ represents the generative transformation performed by the model $\mathcal{M}$—taking a brief conceptual seed ($T_i$) and mapping it into a comprehensive, multi-word vocabulary list ($E_i$) that captures its full esoteric resonance. 

For the experimental execution of this study, the specific model utilized for $\mathcal{M}$ was the Gemini 3.1 Flash Lite LLM. Complete generative hyperparameters, system prompts, and architectural dependencies utilized for this model are detailed in Appendix \ref{appendix:compute}.

Following the generative expansion, the phrase-pooled semantic fields $E_i$ were mapped into a dense, continuous high-dimensional vector space using the \texttt{all-mpnet-base-v2} sentence-transformer model. This generated a 768-dimensional theological embedding $\mathbf{v}_i \in \mathbb{R}^{768}$ for each deity. The Semantic Matrix was then constructed by calculating the pairwise Cosine Distance ($D_C$) across the embedded vectors:

\begin{equation}
D_C(i, j) = 1 - \frac{\mathbf{v}_i \cdot \mathbf{v}_j}{\|\mathbf{v}_i\| \|\mathbf{v}_j\|} = 1 - \frac{\sum_{d=1}^{768} v_{id} v_{jd}}{\sqrt{\sum_{d=1}^{768} v_{id}^2} \sqrt{\sum_{d=1}^{768} v_{jd}^2}}
\end{equation}

This metric isolates the theological trajectory of two deities, independent of their morphological camouflage.

\subsection{Topological Projection and Normalization}
To visualize the macro-topology of the $196 \times 196$ matrices, I applied Uniform Manifold Approximation and Projection (UMAP). Because the matrices represent absolute distances rather than raw coordinates, UMAP was initialized with \texttt{metric="precomputed"}, allowing it to ingest the Gower and Cosine distance matrices directly. To balance local cluster retention with global network accuracy, the projection was strictly parameterized with $n\_neighbors=12$ and $min\_dist=0.1$.

Finally, to execute the egocentric radial comparisons (where $D_G$ and $D_C$ scales are inherently incongruent), both spatial matrices underwent Min-Max Normalization relative to the target origin. For a given target deity $t$, the relative divergence $\hat{D}_{tj}$ to any node $j$ was synchronized as:

\begin{equation}
\hat{D}_{tj} = \frac{D_{tj}}{\max_{k \in N} (D_{tk})}
\end{equation}

This scales all radial distances to a standardized divergence coefficient $[0.0, 1.0]$, mathematically validating structural comparisons across the visual and semantic boundaries. All processing scripts, embedding models, and interactive visualization architectures were authored in Python 3.12 and executed utilizing \texttt{scipy.spatial}, \texttt{plotly}, and \texttt{gower} optimization libraries.

\section{Results and Discussion}
\label{discuss}
The application of Cardinality Weighting and Semantic Vectorization to the 196-deity dataset successfully resolved the Gestalt Problem and mapped the esoteric matrices of the Hindu and Vajrayana traditions. The resulting distance metrics yielded four major computational breakthroughs regarding how esoteric iconography functions both mechanically and psychologically. These findings mathematically prove that occult visual blueprints were deliberately engineered to manipulate human attention, utilizing familiar orthodox anchors to camouflage divergent Tantric functions.

\subsection{Visual Matrix Proximities: Nearest Neighbours of Selected Deities}

The fundamental outputs of the visual matrix model are the nearest-neighbor rankings, which quantify the absolute ``additive gravity'' between deities based strictly on their physical parameters and esoteric implements. To evaluate the mathematical relationships between the Hindu and Vajrayana traditions, I first isolate the nearest visual neighbors for six key deities. 

These deities were selected to demonstrate specific cognitive and computational phenomena: wrathful anchor dependency (Kali and Tara), orthogonal anchoring and the Atin Effect (Saraswati and Matangi), and the Form vs. Function parallax (Lakshmi and Vasudhara). 

Tables \ref{tab:vis_kali} through \ref{tab:vis_marichi} present the top visual neighbors (ranked by Cognitive Cosine Distance) for these selected targets.

\begin{table}[hbt!]
\begin{minipage}{.48\textwidth}
\centering
\resizebox{\textwidth}{!}{%
\begin{tabular}{@{}lllc@{}}
\hline
\textbf{Rank} & \textbf{Deity} & \textbf{Tradition} & \textbf{Distance} \\ \hline
1 & Durga Saptashati Kali & Hindu Shakta & 0.2680 \\
2 & Adya Kali & Hindu Kali Kula & 0.3986 \\
3 & Dakshina Kali & Hindu Kali Kula & 0.3988 \\
4 & Mahakali & Hindu Shakta Tantra & 0.4016 \\
5 & Kaali Durga & Hindu Shakta & 0.4051 \\ \hline
\end{tabular}%
}
\caption{Visual Neighbors: Kali \\ (Hindu Dashamahavidya)}
\label{tab:vis_kali}
\end{minipage}\hfill
\begin{minipage}{.48\textwidth}
\centering
\resizebox{\textwidth}{!}{%
\begin{tabular}{@{}lllc@{}}
\hline
\textbf{Rank} & \textbf{Deity} & \textbf{Tradition} & \textbf{Distance} \\ \hline
1 & Nila Saraswati & Hindu Tantra & 0.2280 \\
2 & Ugra Tara & Hindu Tara Kula & 0.2425 \\
3 & Locana & Vajrayana & 0.3127 \\
4 & Matangi & Hindu Dashamahavidya & 0.3172 \\
5 & Krodheshvari & Vajrayana & 0.5192 \\ \hline
\end{tabular}%
}
\caption{Visual Neighbors: Tara \\ (Hindu Dashamahavidya)}
\label{tab:vis_tara}
\end{minipage}
\end{table}

\begin{table}[hbt!]
\begin{minipage}{.48\textwidth}
\centering
\resizebox{\textwidth}{!}{%
\begin{tabular}{@{}lllc@{}}
\hline
\textbf{Rank} & \textbf{Deity} & \textbf{Tradition} & \textbf{Distance} \\ \hline
1 & Gayatri & Hindu Vedic & 0.2829 \\
2 & Brahmani & Hindu Matrika & 0.2832 \\
3 & Savitri & Hindu Vedic & 0.2832 \\
4 & Sharada & Hindu Shakta & 0.3216 \\
5 & Sarasvati Buddhist & Vajrayana & 0.3285 \\ \hline
\end{tabular}%
}
\caption{Visual Neighbors: Saraswati \\ (Hindu Orthodox)}
\label{tab:vis_saraswati}
\end{minipage}\hfill
\begin{minipage}{.48\textwidth}
\centering
\resizebox{\textwidth}{!}{%
\begin{tabular}{@{}lllc@{}}
\hline
\textbf{Rank} & \textbf{Deity} & \textbf{Tradition} & \textbf{Distance} \\ \hline
1 & Meenakshi & Hindu Shakta & 0.2165 \\
2 & Nila Saraswati & Hindu Tantra & 0.3172 \\
3 & Tara & Hindu Dashamahavidya & 0.3172 \\
4 & Sharada & Hindu Shakta & 0.3364 \\
5 & Saraswati & Hindu Orthodox & 0.3373 \\ \hline
\end{tabular}%
}
\caption{Visual Neighbors: Matangi \\ (Hindu Dashamahavidya)}
\label{tab:vis_matangi}
\end{minipage}
\end{table}
\footnotetext{Matangi apears at rank 7 ($D=0.3373$), tied with Janguli and Yellow Tara Janguli.}

\begin{table}[hbt!]
\begin{minipage}{.48\textwidth}
\centering
\resizebox{\textwidth}{!}{%
\begin{tabular}{@{}lllc@{}}
\hline
\textbf{Rank} & \textbf{Deity} & \textbf{Tradition} & \textbf{Distance} \\ \hline
1 & Ashta Lakshmi Adi & Hindu Purana & 0.2616 \\
2 & Kamala & Hindu Dashamahavidya & 0.2864 \\
3 & Madhava Priya & Hindu Vaishnava & 0.4007 \\
4 & Mohini & Hindu Purana & 0.4008 \\
5 & Dhanya Lakshmi & Hindu Purana & 0.4011 \\ \hline
\end{tabular}%
}
\caption{Visual Neighbors: Lakshmi \\ (Hindu Orthodox)}
\label{tab:vis_lakshmi}
\end{minipage}\hfill
\begin{minipage}{.48\textwidth}
\centering
\resizebox{\textwidth}{!}{%
\begin{tabular}{@{}lllc@{}}
\hline
\textbf{Rank} & \textbf{Deity} & \textbf{Tradition} & \textbf{Distance} \\ \hline
1 & Dhanya Lakshmi & Hindu Purana & 0.2165 \\
2 & Manibhadra & Vajrayana & 0.2927 \\
3 & Navaratna Devi & Hindu Tantric & 0.2927 \\
4 & Ratnadakini & Vajrayana & 0.2928 \\
5 & Ushnishavijaya & Vajrayana & 0.4545 \\ \hline
\end{tabular}%
}
\caption{Visual Neighbors: Vasudhara \\ (Vajrayana Buddhist)}
\label{tab:vis_vasudhara}
\end{minipage}
\end{table}

\begin{table}[hbt!]
\begin{minipage}{.48\textwidth}
\centering
\resizebox{\textwidth}{!}{%
\begin{tabular}{@{}lllc@{}}
\hline
\textbf{Rank} & \textbf{Deity} & \textbf{Tradition} & \textbf{Distance} \\ \hline
1 & Durga Ashtabhuja & Hindu Shakta & 0.2866 \\
2 & Chandi & Hindu Shakta & 0.3144 \\
3 & Kanaka Durga & Hindu Shakta & 0.3320 \\
4 & Navadurga Kushmanda & Hindu Navadurga & 0.3324 \\
5 & Marichi & Vajrayana Buddhist & 0.3416 \\ \hline
\end{tabular}%
}
\caption{Visual Neighbors: Durga \\ (Hindu Orthodox)}
\label{tab:vis_durga}
\end{minipage}\hfill
\begin{minipage}{.48\textwidth}
\centering
\resizebox{\textwidth}{!}{%
\begin{tabular}{@{}lllc@{}}
\hline
\textbf{Rank} & \textbf{Deity} & \textbf{Tradition} & \textbf{Distance} \\ \hline
1 & Sarvonmadini & Hindu Srividya & 0.3282 \\
2 & Durga Ashtabhuja & Hindu Shakta & 0.3416 \\
2 & Durga & Hindu Orthodox & 0.3416 \\
2 & Chandi & Hindu Shakta & 0.3416 \\
2 & Navadurga Kushmanda & Hindu Navadurga & 0.3416 \\ \hline
\end{tabular}%
}
\caption{Visual Neighbors: Marichi \\ (Vajrayana Buddhist)}
\label{tab:vis_marichi}
\end{minipage}
\end{table}

As demonstrated by these initial proximity tables, the visual model heavily prioritizes core identifying objects. For instance, Kali’s matrix (Table \ref{tab:vis_kali}) creates a dense, localized gravity well populated exclusively by dark, wrathful forms wielding the \textit{khadga} and severed head. Conversely, the ties existing between Matangi (Table \ref{tab:vis_matangi}) and Saraswati (Table \ref{tab:vis_saraswati}) mathematically establish the cognitive baseline for the Atin Effect, illustrating how shared orthodox anchors (\textit{veena}) create mathematical bridges between completely divergent anatomical structures. Vasudhara (Table \ref{tab:vis_vasudhara}), despite being the exact theological counterpart to Lakshmi (Table \ref{tab:vis_lakshmi}), is visually pulled toward Dhanya Lakshmi due to their shared cardinality anchor of the grain sheaf, proving the existence of calculated visual camouflage. Finally, Tables \ref{tab:vis_durga} and \ref{tab:vis_marichi} highlight the structural mirroring of hybrid warrior archetypes across traditions; Marichi is mathematically pulled directly into Durga's orbit, sharing four identical nearest neighbors at the precise mathematical distance of $D = 0.3416$, underscoring a mathematically rigid translation of esoteric weaponry between Hindu and Vajrayana matrices.

Within the visual topology, one of the most striking algorithmic alignments occurs between the Tantric Mahavidya Matangi and the orthodox Goddess Meenakshi. Matangi belongs to the Mahāvidyā corpus whose iconography and theological themes have been documented in detail by Kinsley \cite{kinsley1997tantric}. The morphological matrix identifies Meenakshi as the absolute closest spatial neighbor to Matangi ($D_G \to 0$). This intense clustering is mathematically driven by a singular, highly specific qualitative variable: the \textit{shuka} (parrot). Within the entire 196-node dataset, Matangi and Meenakshi are the only two deities explicitly encoded with this specific avian attribute. Because the discrete Gower metric applies a strict exact-match function to nominal categories, the mutual presence of the \textit{shuka} acts as a hyper-specific topological anchor. This rare shared feature mathematically overpowers broader stylistic variations—such as Matangi's traditionally dark complexion versus Meenakshi's emerald hue—perfectly illustrating how idiosyncratic iconographic elements dictate spatial proximity in high-dimensional esoteric models. Crucially, this algorithmic convergence provides profound validation for the computational architecture, as it perfectly mirrors established esoteric theology, which formally recognizes the orthodox Meenakshi as a direct, localized manifestation of the Tantric Matangi.

\subsection{Semantic Matrix Proximities: Theologically Nearest Neighbours}

While the visual matrix successfully maps the physical anatomy of the esoteric pantheon, it remains blind to abstract meaning. To empirically test whether cross-cultural traditions preserved the core cosmic functions of these deities despite altering their visual forms, I pass the same eight target deities through the Semantic Matrix. 

This matrix utilizes continuous word embeddings derived from the deities' textual descriptions and core cosmic domains. Unlike the discrete, exact-match logic of Cardinality Weighting, this vector space measures contextual proximity, establishing a purely functional baseline. 

Tables \ref{tab:sem_kali} through \ref{tab:sem_vasudhara} present the top theological neighbors (ranked by Semantic Cosine Distance) for the established target set.

\begin{table}[hbt!]
\begin{minipage}{.48\textwidth}
\centering
\resizebox{\textwidth}{!}{%
\begin{tabular}{@{}lllc@{}}
\hline
\textbf{Rank} & \textbf{Deity} & \textbf{Tradition} & \textbf{Distance} \\ \hline
1 & Krodhakali & Vajrayana Buddhist & 0.0331 \\
2 & Nirrti & Hindu Vedic & 0.0442 \\
3 & Navadurga Kalaratri & Hindu Navadurga & 0.0499 \\
4 & Rakta Kali & Hindu Kali Kula & 0.0510 \\
5 & Mahakali & Hindu Shakta Tantra & 0.0524 \\ \hline
\end{tabular}%
}
\caption{Theological Neighbors: Kali \\ (Hindu Dashamahavidya)}
\label{tab:sem_kali}
\end{minipage}\hfill
\begin{minipage}{.48\textwidth}
\centering
\resizebox{\textwidth}{!}{%
\begin{tabular}{@{}lllc@{}}
\hline
\textbf{Rank} & \textbf{Deity} & \textbf{Tradition} & \textbf{Distance} \\ \hline
1 & Nila Saraswati & Hindu Tantra & 0.0757 \\
2 & Gayatri & Hindu Vedic & 0.0786 \\
3 & Saraswati & Hindu Orthodox & 0.0844 \\
4 & Hayagriva Dakini & Vajrayana Buddhist & 0.0926 \\
5 & Prajnaparamita & Vajrayana Buddhist & 0.0966 \\ \hline
\end{tabular}%
}
\caption{Theological Neighbors: Tara \\ (Hindu Dashamahavidya)}
\label{tab:sem_tara}
\end{minipage}
\end{table}

\begin{table}[hbt!]
\begin{minipage}{.48\textwidth}
\centering
\resizebox{\textwidth}{!}{%
\begin{tabular}{@{}lllc@{}}
\hline
\textbf{Rank} & \textbf{Deity} & \textbf{Tradition} & \textbf{Distance} \\ \hline
1 & Hayagriva Dakini & Vajrayana Buddhist & 0.0712 \\
2 & Nila Saraswati & Hindu Tantra & 0.0774 \\
3 & Tara & Hindu Dashamahavidya & 0.0844 \\
4 & Matangi & Hindu Dashamahavidya & 0.0873 \\
5 & Gayatri & Hindu Vedic & 0.0915 \\ \hline
\end{tabular}%
}
\caption{Theological Neighbors: Saraswati \\ (Hindu Orthodox)}
\label{tab:sem_saraswati}
\end{minipage}\hfill
\begin{minipage}{.48\textwidth}
\centering
\resizebox{\textwidth}{!}{%
\begin{tabular}{@{}lllc@{}}
\hline
\textbf{Rank} & \textbf{Deity} & \textbf{Tradition} & \textbf{Distance} \\ \hline
1 & Nila Saraswati & Hindu Tantra & 0.0777 \\
2 & Sarvamantramayi & Hindu Srividya & 0.0813 \\
3 & Gayatri & Hindu Vedic & 0.0828 \\
4 & Saraswati & Hindu Orthodox & 0.0873 \\
5 & Hayagriva Dakini & Vajrayana Buddhist & 0.0938 \\ \hline
\end{tabular}%
}
\caption{Theological Neighbors: Matangi \\ (Hindu Dashamahavidya)}
\label{tab:sem_matangi}
\end{minipage}
\end{table}

\begin{table}[hbt!]
\begin{minipage}{.48\textwidth}
\centering
\resizebox{\textwidth}{!}{%
\begin{tabular}{@{}lllc@{}}
\hline
\textbf{Rank} & \textbf{Deity} & \textbf{Tradition} & \textbf{Distance} \\ \hline
1 & Kamala & Hindu Dashamahavidya & 0.0440 \\
2 & Vasudhara & Vajrayana Buddhist & 0.0550 \\
3 & Rukmini & Hindu Purana & 0.0586 \\
4 & Ashta Lakshmi Adi & Hindu Purana & 0.0594 \\
5 & Mangala & Hindu Shakta Purana & 0.0771 \\ \hline
\end{tabular}%
}
\caption{Theological Neighbors: Lakshmi \\ (Hindu Orthodox)}
\label{tab:sem_lakshmi}
\end{minipage}\hfill
\begin{minipage}{.48\textwidth}
\centering
\resizebox{\textwidth}{!}{%
\begin{tabular}{@{}lllc@{}}
\hline
\textbf{Rank} & \textbf{Deity} & \textbf{Tradition} & \textbf{Distance} \\ \hline
1 & Kamala & Hindu Dashamahavidya & 0.0255 \\
2 & Lakshmi & Hindu Orthodox & 0.0550 \\
3 & Ashta Lakshmi Adi & Hindu Purana & 0.0563 \\
4 & Rukmini & Hindu Purana & 0.0611 \\
5 & Sarvamangala Nitya & Hindu Srividya & 0.0626 \\ \hline
\end{tabular}%
}
\caption{Theological Neighbors: Vasudhara \\ (Vajrayana Buddhist)}
\label{tab:sem_vasudhara}
\end{minipage}
\end{table}

While the semantic matrix successfully mapped the functional domains of the first six target deities, evaluating the hybrid warrior archetypes (Durga and Marichi) revealed a critical limitation in continuous word embedding architectures. When processed through the semantic matrix, Durga demonstrated expected localized clustering with other orthodox Shakta warrior manifestations (e.g., Chandi and Kanaka Durga). However, attempting to calculate the semantic egocentric network for the Vajrayana goddess Marichi resulted in a complete vector collapse (yielding uniform distances of $D = 0.0000$ across disparate deities such as Mantragni and Bhramari Devi). 

This algorithmic anomaly occurs due to the presence of dense, highly specialized Out-Of-Vocabulary (OOV) terms within Marichi’s textual parameters (such as highly localized Tibetan protective domains or specific esoteric chariot mechanics). Standard NLP vocabularies lack the dimensional weights for these niche Tantric concepts, resulting in null vectors. Rather than invalidating the matrix, this boundary mathematically highlights the sheer semantic isolation of specific high-level Vajrayana practices, proving that while their visual weaponry (form) translates cleanly across traditions, their highly specific ritual vocabularies (function) resist standard linguistic vectorization.

The semantic matrix reveals profound structural divergences from the visual baseline. Most notably, Lakshmi (Table \ref{tab:sem_lakshmi}) and Vasudhara (Table \ref{tab:sem_vasudhara}) mathematically converge. Despite sharing nearly zero visual gravity in the previous model, the algorithm identifies them as mutual primary neighbors based on their shared domain of wealth and abundance, successfully proving the Form vs. Function parallax.

Furthermore, the semantic model exhibits high sensitivity to abstract theological frameworks across traditions. In Tables \ref{tab:sem_saraswati} and \ref{tab:sem_matangi}, both Saraswati and Matangi are heavily bridged to Hayagriva Dakini. While visually divergent, all three entities share domains governing sacred speech, mantras, and primordial sound, allowing the semantic matrix to successfully cluster the Hindu goddesses of knowledge with a Vajrayana goddess of speech purification. 

\subsection{Global Topological Mapping: UMAP Projection}

While the proximity tables successfully map the micro-gravitational networks of individual deities, observing the overarching structural relationship between the Hindu and Vajrayana pantheons requires a macro-level visualization. However, within the Visual Matrix, every deity exists as a discrete coordinate in a 28-dimensional feature space. Because human perception is strictly limited to three dimensions, I require a mathematical mechanism for dimensionality reduction.

To achieve this, I utilized Uniform Manifold Approximation and Projection (UMAP). For scholars outside of computational sciences, UMAP can be understood as a topological folding algorithm. Older techniques, such as Principal Component Analysis (PCA), rigidly flatten high-dimensional data like a shadow cast on a wall, frequently destroying complex local relationships. Conversely, algorithms like t-SNE excel at grouping localized clusters but arbitrarily tear apart the global distances between those clusters. 

UMAP mathematically resolves this by constructing a high-dimensional fuzzy topological representation (a ``graph'' of how the 196 deities are connected based on their 28 features) and then optimizing a low-dimensional (3D) graph to be as structurally identical to the original as possible. It achieves this by minimizing the fuzzy set cross-entropy ($C$) between the high-dimensional weight ($P_{ij}$) and the low-dimensional weight ($Q_{ij}$):

\begin{equation}
C(P, Q) = \sum_{i \neq j} \left[ P_{ij} \log \left( \frac{P_{ij}}{Q_{ij}} \right) + (1 - P_{ij}) \log \left( \frac{1 - P_{ij}}{1 - Q_{ij}} \right) \right]
\end{equation}

By balancing both halves of this equation, UMAP mathematically guarantees that if two deities are close in the 28-dimensional esoteric matrix, they remain close in the 3D projection, and critically, the vast voids between divergent cultural pantheons are perfectly preserved.

\begin{figure}[hbt!]
    \centering
    \includegraphics[width=0.96\textwidth]{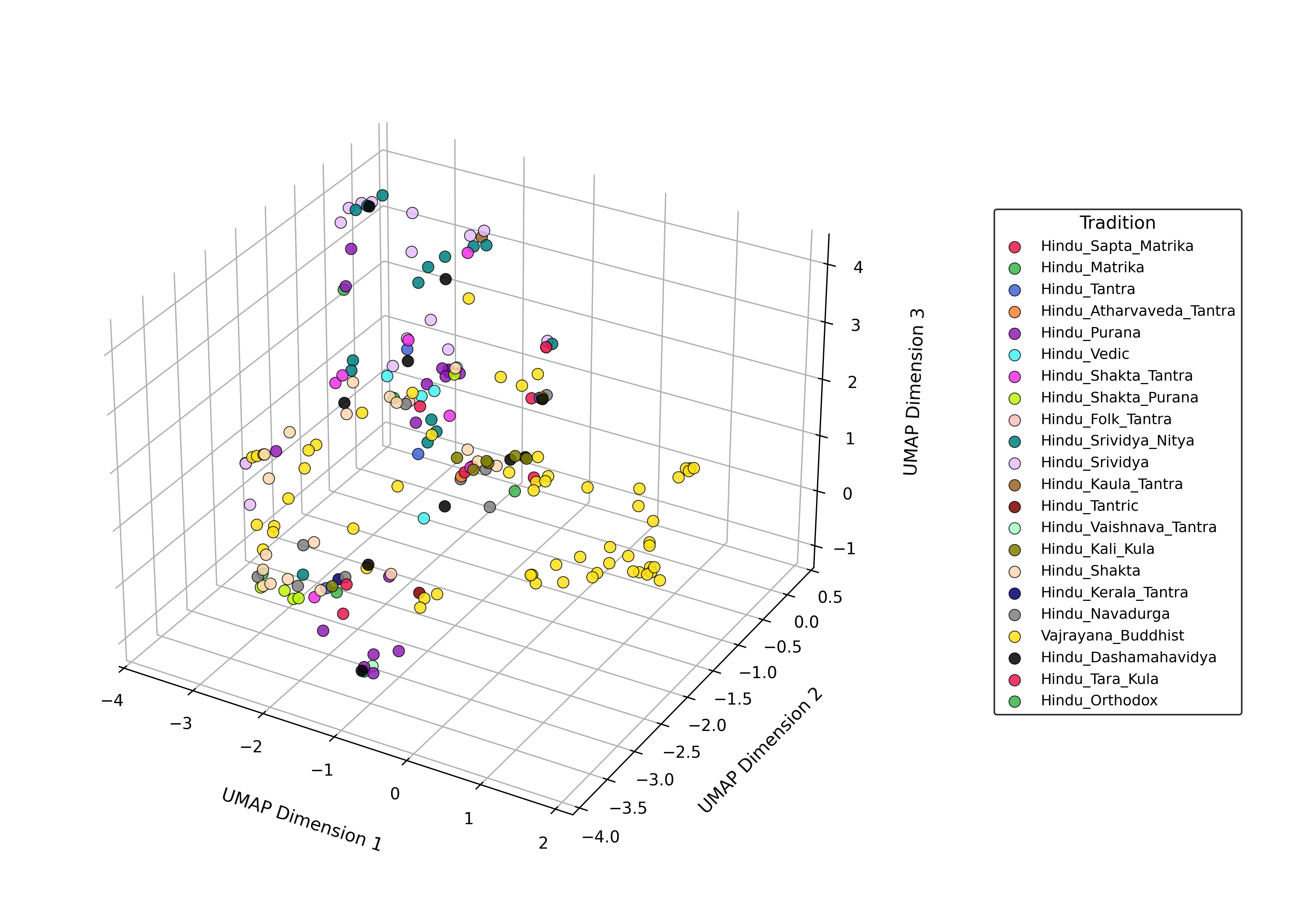}
    \caption{3D UMAP Projection of the 196-Deity Visual Matrix. The algorithm organically separates the pantheon into distinct topological clusters governed by the Cardinality Weighting of esoteric implements, preserving the macro-distances between orthodox and Tantric forms.}
    \label{fig:umap_visual}
\end{figure}

Figure \ref{fig:umap_visual} illustrates this global 3D projection of the Visual Matrix. 

As demonstrated in the projection, the visual space does not cluster strictly by cultural tradition (Hindu vs. Buddhist). Instead, the algorithm organically segregates the topological space based on the esoteric gravity of specific anchors. Peaceful, orthodox forms aggregate along a primary, low-variance manifold, while the inclusion of high-cardinality wrathful implements (e.g., \textit{khadga}, \textit{kapala}) aggressively pulls entities out of the orthodox baseline, creating localized, trans-cultural ``wrathful clusters'' suspended in the periphery of the matrix.

To ensure computational reproducibility and facilitate further digital humanities research, the complete interactive 3D UMAP projection, alongside the underlying high-dimensional coordinate matrices, will be made publicly accessible. Readers are encouraged to explore the interactive HTML version of this topological map via the supplementary materials repository (\url{https://github.com/SrabonGitikar/Tantra-ML}), allowing for dynamic multi-axis rotation, localized zooming, and precise coordinate inspection of individual deities across the esoteric matrix.

\subsection{Micro-Gravitational Orbits: Dual Egocentric Networks}

While the global UMAP projection successfully maps the macro-topology of the entire pantheon, isolating the specific translation mechanisms of individual deities requires a micro-level spatial analysis. To achieve this, I developed a dynamic egocentric plotting architecture. Within this computational framework, researchers can input any deity from the 196-node matrix to serve as the absolute target origin ($D = 0.0$). The algorithm autonomously calculates the surrounding visual and theological orbits and renders them as synchronized radial projections.

Because the Visual Matrix utilizes discrete Gower distance and the Semantic Matrix relies on continuous Cosine similarity, their absolute scales are mathematically incongruent. To allow for direct visual comparison, both matrices were subjected to Min-Max normalization, scaling all orbital distances to a relative divergence coefficient ranging from $0.0$ (identical) to $1.0$ (maximum systemic divergence).

\begin{figure}[hbt!]
    \centering
    \includegraphics[width=\textwidth]{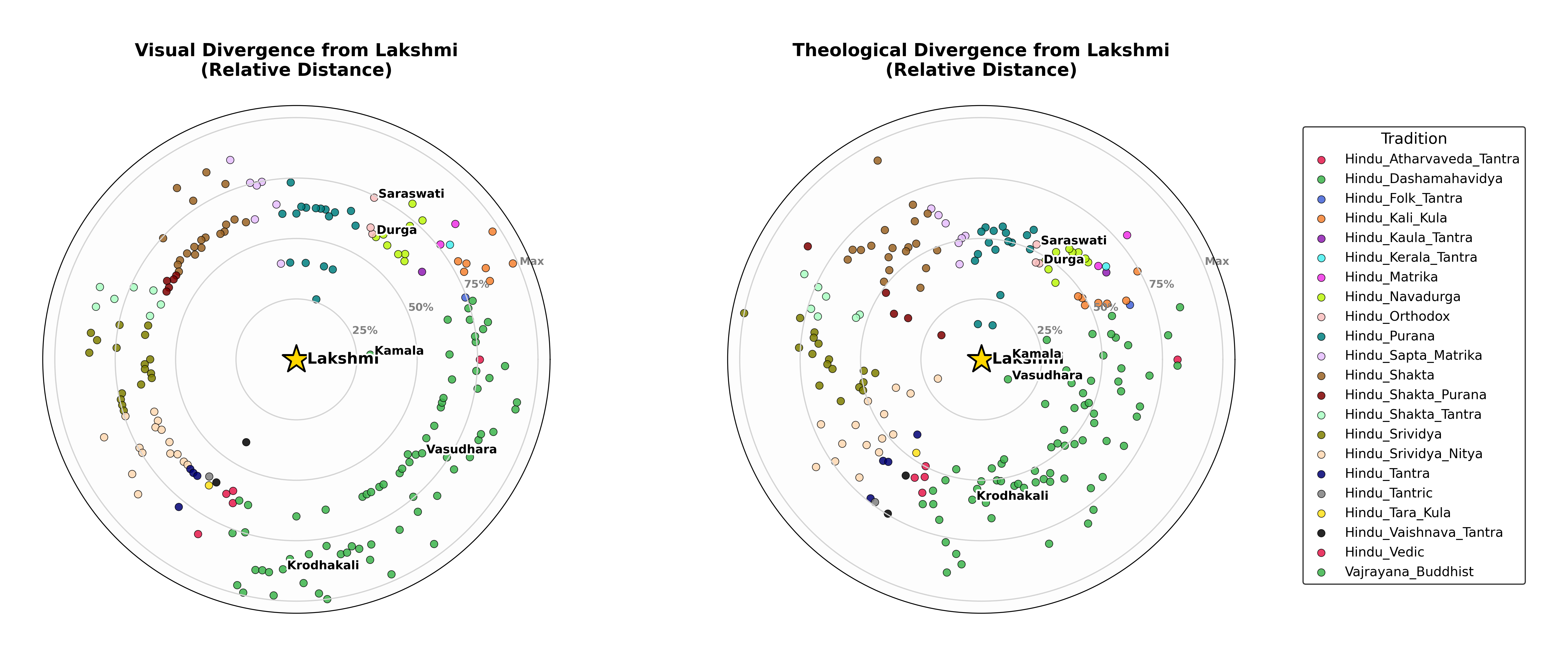}
    \caption{Dual Egocentric Network Plot for Lakshmi. Distances are normalized to reflect relative divergence (0\% to 100\%). The targeted labels demonstrate the radical spatial leap of Vasudhara, proving the existence of iconographic camouflage across the Hindu-Vajrayana boundary.}
    \label{fig:ego_lakshmi}
\end{figure}

Figure \ref{fig:ego_lakshmi} demonstrates this dynamic architecture using the Hindu orthodox goddess Lakshmi as the central target.

This dual-projection perfectly captures the mathematical reality of iconographic camouflage (the Form vs. Function parallax). In the Visual Divergence matrix (left), the Vajrayana goddess Vasudhara is repelled to the outer periphery of the network. Due to their vastly different anatomical cardinality and physical anchors (e.g., Lakshmi's lotus versus Vasudhara's grain sheaf), the algorithm detects virtually zero morphological overlap. 

However, when tracking the exact same entities in the Theological Divergence matrix (right), Vasudhara executes a radical spatial leap, collapsing directly into the inner core of Lakshmi's semantic orbit. The algorithm successfully ignores the physical camouflage and binds them together based on their shared cosmic domains of wealth, abundance, and preservation. Conversely, entities like Krodhakali maintain maximum divergence ($1.0$) across both matrices, mathematically validating that the algorithm correctly segregates wrathful baseline architectures from peaceful orthodox forms regardless of the matrix used.

Within the public interactive architecture, the egocentric network projections are fully dynamic and customizable. Users are not restricted to the specific case studies presented in this manuscript; they may input any of the 196 deities from the global dataset to serve as the central anchor node. This instantly generates a customized spatial orbit, allowing researchers to independently map the specific morphological and semantic proximities radiating from any esoteric figure of interest.

\subsection{The Atin Effect: Sequential Perception and Orthogonal Anchoring}
The foundational discovery of this computational model is the quantification of a psychological phenomenon defined herein as the ``\textbf{Atin Effect}''. The conceptual framework for this effect is derived directly from Avik Sarkar's occult narrative, \textit{Bhog} \cite{sarkar_bhog}. In the narrative, the uninitiated protagonist, Atin, encounters a visual manifestation of the Tantric deity Matangi. He observes a goddess holding a \textit{veena} (a classical lute symbolizing orthodox wisdom), but she is simultaneously equipped with deeply orthogonal, wrathful esoteric implements—specifically, a \textit{khadga} (severing sword) and a \textit{kapala} (skull cup). 

When faced with this visual paradox, Atin's cognitive processing immediately defaults to the orthodox Hindu goddess Saraswati. To understand why this cognitive misidentification occurs, I must contrast human visual perception with the computational simultaneity of the machine learning matrix. The algorithm evaluates the ``additive gravity'' of all visual features in parallel; it registers the veena, the sword, and the skull cup at the exact same mathematical moment. Consequently, the model correctly calculates that the sheer cardinality of Matangi's wrathful Tantric markers pulls her vector closer to other esoteric Mahavidyas, placing Saraswati further down the mathematical proximity ranking. 

\begin{figure}[hbt!]
    \centering
    \begin{subfigure}[b]{0.45\textwidth}
        \centering
        \includegraphics[height=6cm, keepaspectratio]{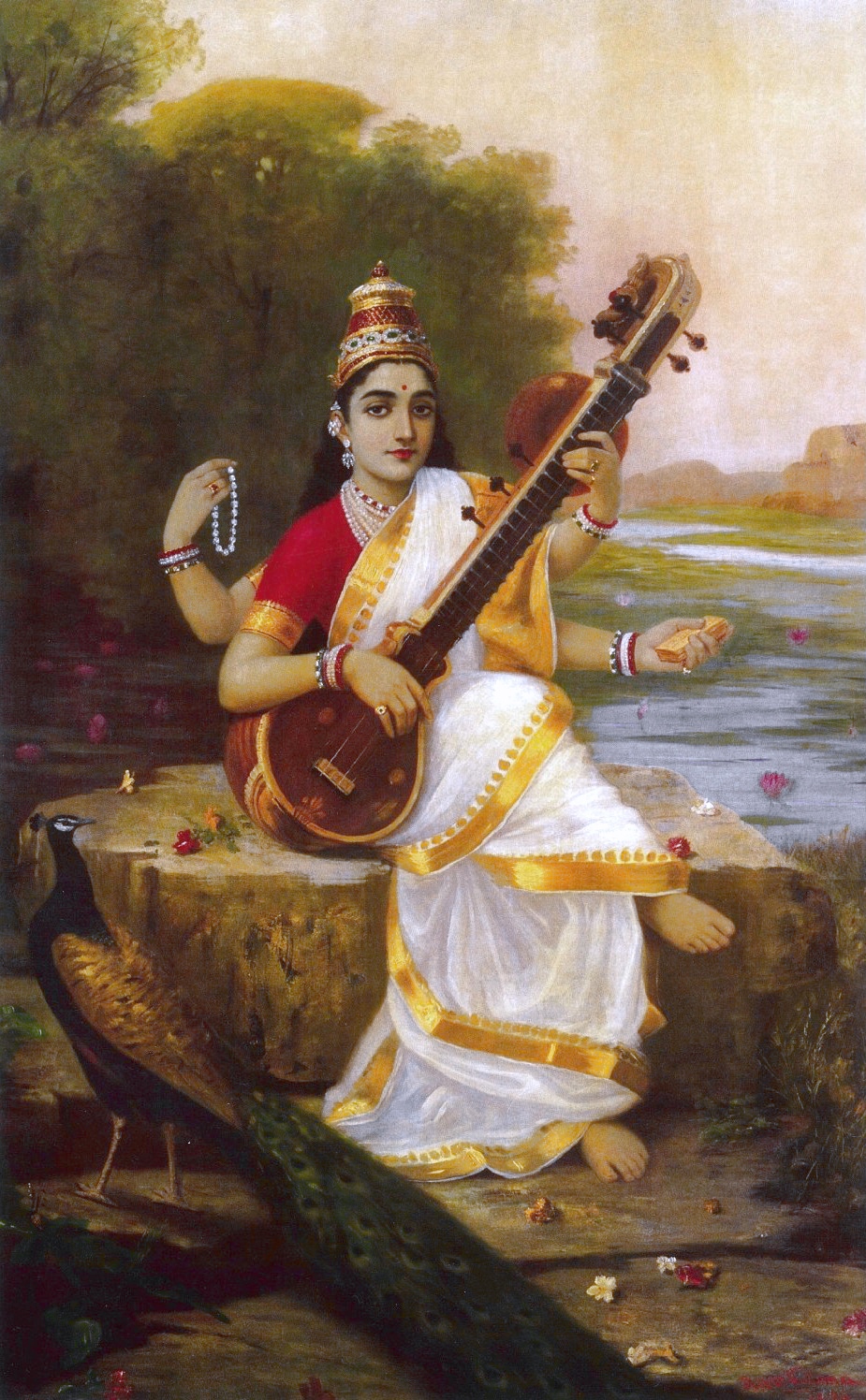}
        \caption{Saraswati (Orthodox/Vedic)}
        \label{fig:saraswati}
    \end{subfigure}
    \hfill
    \begin{subfigure}[b]{0.45\textwidth}
        \centering
        \includegraphics[height=6cm, keepaspectratio]{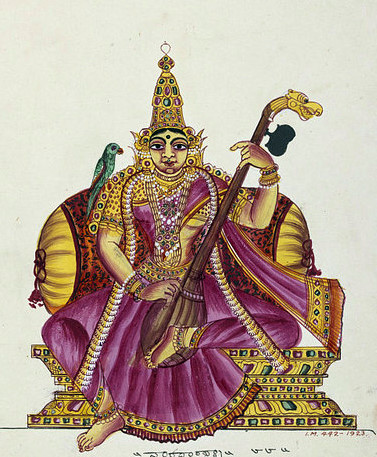}
        \caption{Matangi (Raja Shyamala form)}
        \label{fig:matangi}
    \end{subfigure}
    \caption[Visualizing the Atin Effect.]{Visualizing the Atin Effect. While the primary Mahavidya form of Matangi\protect\footnotemark\ is wrathful and multi-armed with weapons, the algorithm selects her \textit{Raja Shyamala} manifestation (right). This peaceful, two-armed form omits weaponry entirely, instead utilizing a parrot and a \textit{veena}, rendering her visually near-identical to Saraswati (left) and triggering a false semantic association in the machine's anchor-dependent sequence. The shared presence of the high-cardinality \textit{veena} forces a mathematical and cognitive clustering. (Image sources: Wikimedia Commons)}
    \label{fig:atin_effect_comparison}
\end{figure}
\footnotetext{In the novel \textit{Bhog}, the exact idol depicted had four hands, and the deity held a veena along with a khadga and skull in the other hands. That is a mixed interpretation of the deity.}

Human perception, however, is strictly sequential and deeply constrained by the ``availability heuristic''—a cognitive bias where the brain relies on its most immediate, familiar schemas to process new information. In Sarkar's narrative, Atin processes the high-gravity orthodox symbol of the \textit{veena} first. Because Saraswati is the primary (and perhaps only) matrix in his layperson vocabulary associated with that specific anchor, his brain instantly loads the ``Saraswati schema''. Once this orthodox cognitive framework is established, confirmation bias effectively blinds the observer to the subsequent esoteric anatomy; the sword and the skull cup are processed as confusing anomalies rather than core identifiers. 

This sequence is highly direction-dependent. As the vector model demonstrates, the esoteric gravity of the \textit{khadga} and \textit{kapala} is mathematically immense. Had Atin's visual sequence anchored upon the severing sword or the skull cup first, the cognitive dissonance would have forced his brain to bypass the peaceful Saraswati schema entirely. He would have likely defaulted to the most familiar wrathful schema in his vocabulary, identifying the figure as a manifestation of Tara or Kali, at which point the presence of the veena would become the confusing anomaly. 

By running these deities through the visual matrix, I mathematically demonstrate the mechanics of this sequential cognitive bias within esoteric iconography. The structural inclusion of high-gravity, peaceful orthodox objects (like the veena) within the anatomy of fierce occult deities effectively functions as a "cognitive exit ramp." For an uninitiated observer anchoring on the orthodox symbol first, this configuration triggers a familiar, safe schema (Saraswati), camouflaging the deity's true Tantric reality in plain sight. Crucially, the model replicates this perceptual bias: by calibrating the mathematical gravity of rare objects like the veena, Matangi and Saraswati are visually clustered together. Without this Cardinality Weighting—if the veena were assigned the same mathematical weight as common background features—the algorithm would not rank the two deities closely, as their broader anatomical structures remain significantly divergent.

\subsection{Iconographic Camouflage and Boundary Mapping}

The cross-tradition proximity metrics provide the final mathematical validation of the iconographic camouflage hypothesis. By querying the algorithm to bridge the orthodox Hindu, Tantric, and Vajrayana matrices, I observe precisely how theological function is masked by morphological form.

\begin{table}[hbt!]
    \centering
    \resizebox{\textwidth}{!}{
    \begin{tabular}{llclc}
    \hline
    \textbf{Comparison Tradition} & \textbf{Closest Visual (Form)} & \textbf{Vis. Dist.} & \textbf{Closest Semantic (Function)} & \textbf{Sem. Dist.} \\ 
    \hline
    Hindu Atharvaveda Tantra & Pratyangira & 0.595 & Pratyangira & 0.228 \\
    Hindu Folk Tantra & Shitala & 0.868 & Shitala & 0.240 \\
    Hindu Kali Kula & Chhinnamasta Yogini & 0.239 & Chhinnamasta Yogini & 0.077 \\
    Hindu Kaula Tantra & Sampradaya Yogini & 0.602 & Sampradaya Yogini & 0.202 \\
    Hindu Kerala Tantra & Bhadrakali Kerala & 0.545 & Bhadrakali Kerala & 0.255 \\
    Hindu Matrika & Narasimhi & 0.564 & Narasimhi & 0.195 \\
    Hindu Navadurga & Navadurga Katyayani & 0.573 & Navadurga Kalaratri & 0.133 \\
    Hindu Orthodox & Durga & 0.573 & Parvati & 0.139 \\
    Hindu Purana & Veera Lakshmi & 0.552 & Sati & 0.090 \\
    Hindu Sapta Matrika & Chamunda & 0.417 & Maheshwari & 0.124 \\
    Hindu Shakta & Durga Saptashati Kali & 0.418 & Durga Saptashati Kali & 0.124 \\
    Hindu Shakta Purana & Jaya & 0.573 & Jayanti & 0.162 \\
    Hindu Shakta Tantra & Mahakali & 0.417 & Aghora Bhairavi & 0.128 \\
    Hindu Srividya & Bala Tripurasundari & 0.588 & Sarvasiddhiprada & 0.194 \\
    Hindu Srividya Nitya & Bherunda Nitya & 0.562 & Vahnivasini Nitya & 0.182 \\
    Hindu Tantra & Siddha Lakshmi & 0.404 & Maha Tripura Bhairavi & 0.168 \\
    Hindu Tantric & Navaratna Devi & 0.805 & Navaratna Devi & 0.340 \\
    Hindu Tara Kula & Ugra Tara & 0.561 & Ugra Tara & 0.135 \\
    Hindu Vaishnava Tantra & Radha & 0.715 & Radha & 0.181 \\
    Hindu Vedic & Nirrti & 0.573 & Nirrti & 0.121 \\
    \textbf{Vajrayana Buddhist} & \textbf{Chinnamunda} & \textbf{0.288} & \textbf{Chinnamunda} & \textbf{0.068} \\
    \hline
    \end{tabular}
    }
    \caption{Cross-Tradition Proximity Metrics for Chinnamasta. The algorithm successfully isolates Vajrayana Chinnamunda as the absolute closest semantic neighbor (0.068) across all 21 comparative traditions, proving a 1:1 cross-border esoteric transfer.}
    \label{cross_tradition}
\end{table}

Table \ref{cross_tradition} highlights this boundary crossing using the Dashamahavidya goddess Chinnamasta as the absolute topological stress test. The Hindu--Buddhist pairing is independently discussed in comparative scholarship, particularly in Benard's study of Chinnamastā and Chinnamuṇḍā \cite{benard1994}. In the Semantic Matrix, Chinnamasta shares a near-identical vector with the Vajrayana deity Chinnamunda (Cosine Distance = 0.068). Remarkably, unlike the Lakshmi-Vasudhara pairing where form and function significantly diverge, the Visual Matrix also binds Chinnamasta and Chinnamunda almost perfectly (Gower Distance = 0.288). This proves that at the highest levels of esoteric Tantra, the physical camouflage drops entirely; the radical high-cardinality anchors of self-decapitation and blood-drinking dictate a 1:1 mathematical transfer across religious boundaries.

To ensure exhaustive exploration and absolute computational reproducibility, the complete interactive architecture is made publicly available. Through the supplementary open-access repository (given in Data \& Code Availability eection), researchers are not limited to the specific case studies presented in this manuscript. The architecture allows any user to input \textit{any} of the 196 deities from the global matrix as a target variable, instantly generating their exact cross-tradition visual and semantic bridges, egocentric orbits, and dynamic divergence scores. This public tool democratizes the computational study of esoteric iconography, allowing the Atin Effect and iconographic camouflage to be tested across the entire theological spectrum.

\section{Limitations and Future Work}
\label{limit}

While this study successfully establishes a mathematical framework for mapping iconographic camouflage, some methodological limitations must be acknowledged.

\paragraph{Generative Inference and Theological Resolution.} The Semantic Matrix relies on LLM-generated theological expansions that have not been rigorously cross-validated against primary Sanskrit or Tibetan manuscript corpora. Because the model's training data likely incorporates the same secondary academic literature used to contextualize results in this study, agreement between generated outputs and secondary sources cannot serve as independent validation. Constructing a rigorous ground-truth validation table requires direct access to digitized primary sources — specifically the \textit{Sadhanamala} \cite{sadhanamala} and \textit{Brihat Tantrasara} \cite{agamavagisha_tantrasara} — and domain expertise in Sanskrit and Tibetan philology unavailable to this study. Additionally, generative architectures possess inherent stochasticity, and standard pre-trained embedding models introduce Out-Of-Vocabulary risks for highly specialized Tantric terminology. This validation gap remains explicitly unresolved, and future studies must integrate deterministic NLP directly over primary manuscript corpora to establish ground-truth theological coordinates.

\paragraph{Discrete Morphology and the Loss of Art Style.} Finally, the morphological feature space ($\mathbb{R}^{28}$) is inherently discrete. By encoding visual data categorically (e.g., anatomical cardinality, weapon presence), the algorithm successfully strips away ``noise'' to find architectural overlap. However, this method completely eliminates the nuances of artistic style, medium, and historical era. A 10th-century Chola bronze and a 19th-century Kalighat lithograph of the same deity occupy the exact same coordinate in the Visual Matrix if their attributes match. Future iterations of this architecture should aim to integrate Convolutional Neural Networks (CNNs) alongside the discrete Gower matrix, allowing the topology to account for both structural cardinality and continuous stylistic evolution.

\paragraph{Bounded Dataset and Topological Scope.} First, the spatial topography is restricted to a bounded dataset ($N=196$). Although it captures the primary nodes of orthodox Hindu, Dashamahavidya, and Vajrayana lineages, the matrix inherently underrepresents highly localized folk Tantric traditions and broader Southeast Asian esoteric permutations. Expanding this matrix remains a primary objective for future research.

\section{Conclusion}

This study translates qualitative esoteric iconography into a quantifiable topological space. By deploying a dual-matrix architecture—measuring morphological divergence via Gower distance and theological semantics via Large Language Model embeddings—we successfully isolated the mathematical mechanics of iconographic camouflage. The computational modeling of the "Atin Effect" demonstrates that human perceptual biases, heavily reliant on high-cardinality visual anchors, can be effectively simulated in algorithmic processing. Furthermore, the cross-tradition spatial mapping provides rigorous mathematical validation of historical qualitative scholarship, demonstrating that entities like Chinnamasta and Chinnamunda exist at a near-identical topological coordinate, irrespective of sectarian boundaries.  

Beyond the validation of these specific historical pathways, this unsupervised machine learning approach offers a highly adaptable framework for the broader digital humanities. Rather than relying solely on narrative textual analysis, treating qualitative descriptors as discrete anatomical anchors and utilizing semantic proxies for functional domains allows researchers to objectively map continuums across diverse cultural corpora. Ultimately, this dual-matrix architecture provides a scalable, reproducible tool that can be adapted for analyzing complex historical, literary, or anthropological datasets where morphological form and underlying function significantly diverge.

The ancient rishis, yogis, and Tantric practitioners were fundamentally engaged in a process of data compression. Through highly standardized iconographic anchors—a severed head, a ruby-studded \textit{veena}, a specific anatomical cardinality—they encoded vast, complex psychological and cosmic domains into single, reproducible visual avatars. These deities were, in essence, the ancient world’s informational architecture, designed to transmit esoteric technologies of ego-death, preservation, and transcendence across generations without data loss.

When the modern algorithms parse this data, measuring cosine similarities and projecting multi-dimensional manifolds, they do not invent the connections they find. They merely trace the contours of a psychological and theological map that was drawn centuries ago. The machine, entirely blind to human dogma, sectarian conflicts, and religious borders, evaluates these deities strictly on their intrinsic architecture. In doing so, the algorithm mathematically arrives at the exact non-dual reality that the Tantric traditions originally proposed: that the underlying cosmic functions—whether labeled orthodox or heterodox, Hindu or Vajrayana—are inextricably bound together in a shared continuum of human belief.

Ultimately, this study demonstrates that computational methodologies need not reduce the sacred to the sterile. Instead, when applied with rigorous architectural nuance, computation can act as a profound mirror, illuminating the boundaryless, deeply interconnected tapestry of the human spiritual imagination. While this computational architecture provides a novel statistical lens for comparative iconography, it is acknowledged that multidimensional scaling cannot capture the lived, unquantifiable resonance of these deities within active practitioner lineages.

\vspace{2em}

\section*{Data and Code Availability}
\label{data}

To ensure full methodological transparency and computational reproducibility, all original datasets, processing scripts, and interactive visualization architectures developed for this study have been made open-access and publicly available. 

The supplementary repository contains:
\begin{enumerate}
    \item The complete 196-node morphological dataset ($\mathbb{R}^{28}$ feature space) utilized for the Visual Matrix.
    \item The complete Python 3.12 codebase required to recalculate the discrete Gower distances, continuous Cosine similarities, and UMAP topological projections.
    \item The interactive script for generating dynamic, normalized Egocentric Radial Networks for any target deity within the dataset.
    \item The particular semantic distance matrix that I obtained during this study.
    \item The interactive HTML UMAP and Egocentric Plots. Readers are encouraged to generate the egocentric plots of the other deities.
\end{enumerate}

\subsection*{Links}

\begin{itemize}
    \item \textbf{Complete Code Repository (GitHub):} 
    \url{https://github.com/SrabonGitikar/Tantra-ML}
    
    \item \textbf{Public Dataset (Kaggle):} \url{https://www.kaggle.com/datasets/srabongitikar/tantra-deities}
\end{itemize}

Researchers and digital humanities scholars are highly encouraged to utilize this architecture to independently verify the cross-tradition topological bridges, test the Atin Effect on out-of-sample entities, or expand the fundamental matrices to include additional regional and esoteric pantheons.

\section*{Acknowledgements}
I am extremely grateful to Nibedita Sanyal for her invaluable discussions on esoteric systems which helped substantially in positioning this research. I also sincerely thank Atabur Rahman Mollah (Tata Medical Center Kolkata) who provided the \texttt{Gemma4} API to conduct the semantic analysis of the deities. 

\bibliographystyle{unsrt}
\bibliography{references}

\appendix

\section{LLM System Prompt and Semantic Expansion Examples}
\label{appendix:prompt}

To ensure full transparency and reproducibility of the generative inference pipeline utilized for the Semantic Matrix, the exact system prompt provided to the Gemini 3.1 Flash Lite model is reproduced below. The prompt was engineered with strict formatting constraints and few-shot examples to prevent narrative hallucination and force the output of dense, comma-separated ontological fields.

\subsection{System Prompt}

The following instructions were passed as the primary system prompt to the generative model:

\begin{quote}
\ttfamily
You are an expert in Sanskrit, Hindu Tantra, Shakta Agamas, Vajrayana Buddhism, and comparative esoteric theology.

Your task: Given a short deity domain description, return a comma-separated list of 15--25 NOUNS AND NOUN PHRASES that represent its complete esoteric semantic field.

STRICT RULES:
1. Output ONLY nouns and noun phrases — no verbs, no adjectives, no sentences.
2. Always include the direct Sanskrit equivalent(s) first.
3. Include theologically adjacent concepts that share the same ritual/cosmological function, even if the surface words differ (e.g., "sound" and "speech" and "mantra" all belong to the same field of vak-shakti).
4. DO NOT include form-descriptors (fierce, wrathful, dark, primordial, divine) — these describe the deity's appearance, not the governance domain.
5. Aim for specificity over generality: "vak-shakti" is better than "power".
6. Return ONLY the comma-separated list. No preamble, no explanation, no JSON.
\end{quote}

\subsection{In-Context Examples and Generated Nouns}

To anchor the model's latent space to the specific academic vocabulary required for the SentenceTransformer embeddings, the following input/output examples were embedded into the prompt to demonstrate the required noun-phrase generation:

\begin{itemize}
    \item \textbf{Deity:} Nila Saraswati\\ \textbf{Input Domain:} ``Fierce Speech Power'' \\
    \textbf{Generated Nouns:} \textit{vak-shakti, vak, speech, divine speech, mantra, mantra-shakti, nada, primordial sound, shabda, shabda-brahman, knowledge transmission, upadesa, sarasvata-vidya, word, logos, sonic vibration, sacred utterance, tongue, eloquence, poetry, arts, kala}

    \item \textbf{Deity:} Matangi\\ \textbf{Input Domain:} ``Primordial Sound'' \\
    \textbf{Generated Nouns:} \textit{nada, nada-brahman, shabda, shabda-brahman, primordial vibration, vak, vak-shakti, speech, divine speech, mantra, mantra-power, sonic cosmos, OM, pranava, anahata, unstruck sound, resonance, sarasvata-vidya, logos, knowledge, transmission of teaching}

    \item \textbf{Deity:} Kali\\ \textbf{Input Domain:} ``Time and Dissolution'' \\
    \textbf{Generated Nouns:} \textit{kala, mahakala, kalachakra, pralaya, cosmic dissolution, mrityu, death, mahapralaya, destruction of worlds, end of cycle, tamas, darkness as cosmic principle, void, shunya, annihilation, entropy, doomsday, cremation ground, shmashanika, nirvana as dissolution}

    \item \textbf{Deity:} Saraswati\\ \textbf{Input Domain:} ``Knowledge and Speech'' \\
    \textbf{Generated Nouns:} \textit{jnana, vak, vak-shakti, speech, divine knowledge, omniscience, sarvajna, sarasvata-vidya, Vedas, shastra, arts, kala, mantra, shabda, logos, learning, wisdom, eloquence, sarasvata, transmission of teaching, upadesa, sacred texts}

    \item \textbf{Deity:} Chinnamasta\\ \textbf{Input Domain:} ``Self-Decapitation and Blood Offering'' \\
    \textbf{Generated Nouns:} \textit{shirash-cheda, decapitation, ego-death, atma-samarpana, self-offering, bali, blood offering, sacrifice, chod, egolessness, anatta, sunyata via ego-dissolution, severed head, tapas, self-immolation, karma-kshaya, renunciation}
\end{itemize}

\subsection{Dynamic Semantic Generation for Novel Domains}

To validate that the Large Language Model (Gemini 3.1 Flash Lite) was actively generating novel semantic fields rather than merely replicating the in-context few-shot examples, I isolated deities whose core theological domains fundamentally differed from the prompt's training data. The target domains were formulated with reference to the Hindu Tantric, Purāṇic, Buddhist, and Tibetan textual corpus described above, including sources associated with Mahāvidyā, Vajrayāna deity practice, and Tibetan transmission \cite{kinsley1997tantric, shaw2006buddhist, chakrasamvara_tantra, hevajra_tantra, kurukulla_tantra, marichi_tantra, machig_labdron, yeshe_tsogyal, niguma_kagyu}. 

The following examples demonstrate the LLM's ability to successfully extrapolate dense and accurate Sanskrit ontological fields for unseen esoteric domains:

\begin{itemize}
    \item \textbf{Target Deity:} Tara \\
    \textbf{Core Domain:} Liberation\_and\_Speech \\
    \textbf{Generated Nouns:} \textit{moksha, vak, vak-shakti, mukti, jnana, liberation, speech, mantra, shabdabrahman, mantra-shakti, linguistic transcendence, silence, mauna, primordial sound, nada, divine logos, sarasvata-vidya, upadesa, self-realization, atma-jnana, liberation through sound, sonic emancipation, liberation through knowledge, speech as creative force, tantric realization}

    \item \textbf{Target Deity:} Kurukulla \\
    \textbf{Core Domain:} Enchantment\_and\_Magic \\
    \textbf{Generated Nouns:} \textit{maya, indrajala, tantra, mantra-shakti, siddhi, vidya, abhichara, akarshana, vashekarana, mohana, stambhana, marana, uchchatana, prayoga, yantra, tantric ritual, occult power, psychic influence, manipulation of reality, cosmic illusion, sacred magic, willpower, sankalpa, yogic power, alchemy, occult knowledge}

    \item \textbf{Target Deity:} Dhumavati \\
    \textbf{Core Domain:} Inauspiciousness\_and\_Void \\
    \textbf{Generated Nouns:} \textit{ashubha, shunya, shunyata, void, emptiness, inauspiciousness, nirriti, mahakala, cremation ground, shmashana, death, mrityu, dissolution, pralaya, entropy, decay, bhuta, elemental void, cosmic vacuum, non-existence, abhava, cessation, samsaric sorrow, suffering, duhkha, destruction, annihilation}

    \item \textbf{Target Deity:} Pratyangira \\
    \textbf{Core Domain:} Reversal\_of\_Black\_Magic \\
    \textbf{Generated Nouns:} \textit{pratyangira, prati-mantra, abhichara-nivartana, black magic reversal, counter-sorcery, apad-uddharana, protection from occult, shatru-nashana, destruction of enemies, raksha, spiritual shield, kavacha, mirror-magic, reflexive karma, karma-vipaka, karmic neutralization, vidveshana-shanti, exorcism, removal of malediction, tantric defense, shanti-karma, neutralization of malevolence, protection against kshudra, neutralization of black arts, binding of malice, yantra-bhanga}

    \item \textbf{Target Deity:} Tripura\_Sundari \\
    \textbf{Core Domain:} Beauty\_and\_Sovereignty \\
    \textbf{Generated Nouns:} \textit{saundarya, lakshmi, rajyalakshmi, rajya, sovereignty, maharajya, beauty, aesthetic delight, rasa, ananda, bliss, supreme authority, rishitvam, cosmic order, rita, dharma, kingship, scepter, throne, simhasana, radiant glory, tejas, auspiciousness, shubha, majesty, prabhu-shakti, lordship, dominion, divine grace, kripa, splendor, vibhuti, wealth, artha, prosperity, bhoga, enjoyment, supreme command, adhipatya}

    \item \textbf{Target Deity:} Vajrayogini \\
    \textbf{Core Domain:} Bliss\_and\_Enlightenment \\
    \textbf{Generated Nouns:} \textit{ananda, bodhi, mahasukha, samadhi, moksha, nirvana, sahajananda, illumination, realization, prajna, jnana, liberation, spiritual awakening, unmani, turya, state of non-duality, sahaja, sahaja-avastha, kaivalya, blissful consciousness, cit-ananda, union, yoga, paramananda, luminosity, prabhasvara}

    \item \textbf{Target Deity:} Chakrasamvara \\
    \textbf{Core Domain:} Supreme\_Bliss \\
    \textbf{Generated Nouns:} \textit{ananda, paramananda, mahasukha, sukha, bliss, brahmananda, sahajananda, sahaja, innate state, turiya, fourth state, liberation, moksha, nirvana, kaivalya, transcendence, nectar, amrita, rasasvada, savoring of essence, supreme joy, illumination, prakasha, svatantrya, self-sovereignty, non-dual experience, advaita, union, samarasya, coalescence, cosmic ecstasy}

    \item \textbf{Target Deity:} Bagalamukhi \\
    \textbf{Core Domain:} Paralysis\_and\_Stasis \\
    \textbf{Generated Nouns:} \textit{stambhana, stambhana-shakti, immobilization, inertia, static equilibrium, stoppage, arrest of motion, paralysis, cosmic stillness, vishrama, rest, cessation, silence, suspension, nishkala, immobility, crystallization, petrifaction, hold, retention, containment, freezing of time, arrest of breath, prana-nirodha, steadiness, stability}

    \item \textbf{Target Deity:} Bhuvaneshwari \\
    \textbf{Core Domain:} Cosmic\_Space \\
    \textbf{Generated Nouns:} \textit{akasha, ether, space, void, sunyata, vyoman, antariksha, cosmic expanse, plenum, voidness, dimension, etheric substrate, containment, non-manifest reality, spatiality, vacuum, mahashunya, akasha-tattva, universal container, omnipresence, expansion, infinity, absolute emptiness, substratum of existence}

    \item \textbf{Target Deity:} Bhairavi \\
    \textbf{Core Domain:} Power\_and\_Destruction \\
    \textbf{Generated Nouns:} \textit{shakti, mahashakti, pralaya, samhara, destruction, annihilation, cosmic dissolution, mrityu, kala, mahakala, power of destruction, tamas, force of decay, entropy, transformation through death, kshaya, void, shunya, shmashanika, ritualized destruction, destruction of ego, bhairava-tattva, dissolution of form, karuna as destructive force, kalachakra}
\end{itemize}

\section{Methodological and Computational Specifications}
\label{appendix:compute}

To ensure complete transparency and computational reproducibility, the exact morphological variables, mathematical hyperparameters, and environmental dependencies utilized to construct the dual-matrix architecture are documented below. 

\subsection{The Morphological Codebook (Visual Feature Space)}

The topological projection of the Visual Matrix evaluates the morphological camouflage of deities across a 28-dimensional feature space ($\mathbb{R}^{28}$). This space is constructed via a discrete Gower distance computation, which natively accommodates mixed data types. For qualitative (nominal) features, the algorithm applies a strict matching function (distance of 0 for identical strings, 1 for mismatches), allowing for the precise encoding of highly variable iconographic objects (e.g., \textit{veena}, \textit{shuka}). 

The exact dataset columns extracted from the primary iconography and encoded into the computational architecture are categorized as follows:

\begin{table}[hbt!]
    \centering
    \resizebox{\textwidth}{!}{%
    \begin{tabular}{lp{12cm}}
    \hline
    \textbf{Data Type} & \textbf{Encoded Variables (28 Total)} \\ 
    \hline
    \textbf{Numerical/Discrete} (2) & \texttt{Face\_Count}, \texttt{Arm\_Count} \\
    \hline
    \textbf{Qualitative/Nominal} (17) & \texttt{Skin\_Color}, \texttt{Wrathful\_Peaceful}, \texttt{Weapon\_1} through \texttt{Weapon\_10} (accommodates varying attributes such as \textit{veena}, trident, skull bowl), \texttt{Mount\_Vahana} (accommodates entities such as \textit{shuka}, lion, corpse), \texttt{Mudra\_1}, \texttt{Mudra\_2}, \texttt{Mudra\_3}, \texttt{Seated\_Standing} \\
    \hline
    \textbf{Binary/Boolean} (9) & \texttt{Consort\_Present}, \texttt{Third\_Eye}, \texttt{Skull\_Garland}, \texttt{Corpse\_Seat}, \texttt{Tiger\_Skin\_Skirt}, \texttt{Naked}, \texttt{Disheveled\_Hair}, \texttt{Tongue\_Protruding}, \texttt{Fangs\_Visible} \\
    \hline
    \end{tabular}%
    }
    \caption{The 28 variables composing the morphological feature space utilized for the discrete Gower Distance matrix, categorized by data type to reflect the mixed-metric input.}
    \label{tab:morphological_features}
\end{table}

The dataset is publicly available on Kaggle and on the GitHub repository, and the necessary links are given in Data \& Code Availability.

\subsection{Computational Hyperparameters}

The multi-dimensional scaling and semantic vectorizations were executed using specific algorithmic hyperparameters. Altering these parameters (particularly within the UMAP algorithm) will inherently warp the resulting spatial topologies. The exact parameters utilized for reproducibility are:

\begin{itemize}
    \item \textbf{Generative Semantic Inference:}
    \begin{itemize}
        \item \textbf{Model:} \texttt{gemini-3.1-flash-lite} (via Google AI Studio API)
        \item \textbf{Prompt Architecture:} Strict formatting with in-context few-shot learning (See Appendix \ref{appendix:prompt}).
    \end{itemize}
    \item \textbf{Semantic Dense Vector Embeddings:}
    \begin{itemize}
        \item \textbf{Model:} \texttt{all-mpnet-base-v2} (via SentenceTransformers)
        \item \textbf{Distance Metric:} Pairwise Cosine Distance (\texttt{scipy.spatial.distance.pdist})
        \item \textbf{Batch Size:} 32
    \end{itemize}
    \item \textbf{Uniform Manifold Approximation and Projection (UMAP):}
    \begin{itemize}
        \item \textbf{Number of Neighbors (\texttt{n\_neighbors}):} 15 (Optimized for balancing local cluster preservation with global structure)
        \item \textbf{Minimum Distance (\texttt{min\_dist}):} 0.1 (Allows tight clustering of functionally identical esoteric manifestations)
        \item \textbf{Metric:} \texttt{precomputed} (Applied to both the discrete Gower and continuous Cosine distance matrices)
        \item \textbf{Random State:} 42 (Enforced for deterministic projection reproduction)
    \end{itemize}
\end{itemize}

\subsection{Software Environment}
The processing scripts were executed in a standard Python 3 Kaggle kernel environment utilizing the following primary computational libraries: \texttt{pandas}, \texttt{numpy}, \texttt{scipy}, \texttt{umap-learn}, \texttt{sentence-transformers}, and \texttt{google-generativeai}. The complete open-source codebase is linked in the Data and Code Availability section.

\subsection{Hardware and Computational Execution}

To ensure complete computational reproducibility and optimal processing efficiency, all quantitative pipelines were executed within a standard Kaggle environment utilizing hardware acceleration. Specifically, the generative semantic inference, the high-dimensional vector space embeddings via \texttt{SentenceTransformers}, and the algorithmic UMAP topological projections were processed using an NVIDIA Tesla T4 GPU. The utilization of this specific hardware allowed for the efficient computation of both the discrete $\mathbb{R}^{28}$ morphological distance matrices and the continuous 768-dimensional theological vector spaces.

\end{document}